\documentclass{article}
\usepackage{fullpage}
\usepackage[utf8]{inputenc}
\usepackage{amsmath}
\usepackage{amsfonts}
\usepackage{amssymb}
\usepackage{pgf}
\usepackage{tikz}
\usepackage[english,ngerman]{babel}
\usetikzlibrary{arrows,automata,snakes}

\newcommand{\abs}[1]{\left| #1\right|}

\newcommand{\set}[1]{\left\{ #1\right\}}

\newcommand{\wrt}{w.r.t. }

\newcommand{\realrange}[2]{\left[#1, #2\right]}

\newcommand{\unitrange}[2]{\realrange{0}{1}}

\newcommand{\llabel}[1]{\label{\labelprefix:#1}}
\newcommand{\labelprefix}{} 

\newcommand{\discussionsize}{\small}

\newcommand{\notiz}[1]{}
\newcommand{\frage}[1]{}

\newenvironment{code}{\noindent
\begin{tabbing}%
\hspace{2em}\=\hspace{2em}\=\hspace{2em}\=\hspace{2em}\=\hspace{2em}\=%
\hspace{2em}\=\hspace{2em}\=\hspace{2em}\=\hspace{2em}\=\hspace{2em}\=%
\kill}{\end{tabbing}}

\newcommand{\labelcommand}{}
\newcommand{\captiontext}{}
\newsavebox{\codeparam}
\newcounter{lineNumber}
\newenvironment{disscodepos}[3]{%
\renewcommand{\labelcommand}{#2}%
\renewcommand{\captiontext}{#3}%
\sbox{\codeparam}{\parbox{\textwidth}{#3}}%
\begin{figure}[#1]\begin{center}\begin{code}\setcounter{lineNumber}{1}}{%
\end{code}\end{center}\caption{\llabel{\labelcommand}\captiontext}\end{figure}}

{\end{disscodepos}}

\newdimen\endofsize\endofsize=0.5em
\def\endofbeweis{~\quad\hglue\hsize minus\hsize
                 \hbox{\vrule height \endofsize width
\endofsize}\par}

\newcommand{\ignore}[1]{}

\renewcommand{\frage}[1]{[{\sf#1}]\marginpar[\hfill$\Longrightarrow$]{
$\Longleftarrow$}}

\newtheorem{dfn}{Definition}

\title{Fast Detour Computation for Ride Sharing}
\author{Robert Geisberger, Dennis Luxen, Sabine Neubauer,\\ Peter Sanders, Lars Volker\\\normalsize
Universit\"at Karlsruhe (TH), 76128 Karlsruhe, Germany\\\normalsize {\tt
\{geisberger,luxen,sanders\}@ira.uka.de; bine.ka@gmx.de; lv@lekv.de }}
\begin{document}
\selectlanguage{english}
\maketitle
\begin{abstract}
Today's ride sharing services still mimic a better billboard.
They list the offers and allow to search for the source and target city, sometimes enriched with radial search.
So finding a connection between big cities is quite easy. These places are on a list of designated origin and destination points.
But when you want to go from a small town to another small town, even when they are next to a freeway, you run into problems.
You can't find offers that would or could pass by the town easily with little or no detour.
We solve this interesting problem by presenting a fast algorithm that computes the offers with the smallest detours \wrt a request.
Our experiments show that the problem is efficiently solvable in times suitable for a web service implementation. For realistic database size we achieve look-up times of about $5ms$ and a matching rate of $90\%$ instead of just $70\%$ for the simple matching algorithms used today.
\end{abstract}
\section{Introduction}
The concept of ride sharing can be described by the following observation: Two people, that we call driver and passenger, wish to travel from individual starting locations to destinations. A lot of these independent journeys have starting and ending locations that are relatively close to each other. So, for economic reasons the travelers team up for some part of their journeys. They share the same vehicle for some time. Ride sharing creates a trade-off situation for the participants. The cost of driving and owning a vehicle versus the time, money and resources needed to organize a shared ride and then split the overall cost among the participants.

There exist a number of web sites that offer ride sharing matching services to their customers. Unfortunately, most of them suffer from some limitations.

Only a very small and limited subset of all the possible locations is actually modelled. This rather limited modelling has several shortcomings. For customers from sparsely populated areas, it can be quite difficult to attribute one of the selected origin and destination places to their location. Radial search around large cities has been introduced to help selecting approximate start and end points of a trip. The selection has to be done more or less manually because sometimes  radii intersect each other. Also, from a technical point of view, a trip changes it start and end points when it is mapped into a predefined set of only a few locations. As a consequence, a correct ranking of possible matches by detour is too much to expect from such a ride sharing matching system.

Another downside is that matching services do not support what we call lazy pickup. The systems ignore any possible intermediate stop, if they are not given beforehand. Note that equally short routes can take arbitrarily different paths. Consider the following example to visualize the problem. Anne and Bob both live in Germany. Anne, the driver, is from Karlsruhe and wants to go to Berlin. Bob on the other side lives in Frankfurt and would like to travel to Leipzig. Taking the shortest route in our example, Anne drives from Karlsruhe via Nürnberg to Berlin and is never getting close enough to team up with Bob. However, there is a path from Karlsruhe to Berlin via Frankfurt, that also passes by the city of Leipzig and is only about one percent longer than the shortest path. In toady's services it is mandatory to predetermine any possible stops artificially by hand, if they would like to pick up any passengers along a single predetermined route. Obviously, this reduces flexibility a lot. Matches that would have been perfect from a practical point of view, as in our example, are now impossible to make since the route of the trip had to be fixed before it even started.

\section{Our Approach to Ride Sharing}
For many services an offer only fits a request iff their start and destination locations and the possibly prefixed route are identical.
We call such a situation a \textit{perfect fit}.
Some services additionally offer radial search and prefixing of the route.
The existence of these additions shows the demand for better route matching techniques that allow a small detour and intermediate stops, we call that a \textit{reasonable fit}.
However, previous approaches obviously used only features of the database systems they had available to compute the perfect fits.
And we showed in the previous section that the previous approaches are not flexible, miss interesting matches and require a lot of manual work.

We present an algorithmic solution to the situation at hand that leads to better results independently of the users level of cooperation or available database systems.
For that we lift the restriction of a limited set of origin and destination points.
Unfortunately, the probability of perfect fits is close to zero in this setting.
But since we want to compute reasonable fits, our approach considers intermediate stops where driver and passenger might meet and depart later on.
More precisely, we adjust the drivers route to pick up the passenger by allowing an acceptable detour.

We model the road network as a Graph $G=(V,E)$. A Path $P$ is a series of nodes $P=\langle v_1, \ldots, v_2\rangle\in V$ with $(v_i,v_{i+1})\in E$. The length $c(P)$ is the sum of the weights of all edges in $P$. Furthermore, $\mu(u,v)$ denotes the length of a shortest path in $G$ for $u,v\in V$.
\begin{dfn}
We say that an offer $o=(s,t,\epsilon)$ and a request $g=(s', t')$ form a reasonable fit iff there exists a Path $P=\langle u,\ldots,v,\ldots,w,\ldots,x\rangle$ in G with $c(P)\leq(1+\epsilon)\cdot\mu(s,t)$ and $u,v \in \{s,s'\}, u\not=v$ and $w,x\in \{t,t'\}, w\not=x$.
\end{dfn}
The \textit{detour} of any participant is calculated as the ratio between the minimal additional distance necessary to match request and offer and the length of shortest path while traveling without any detour. For a scenario where a driver uses a car an a passenger might not, we propose to weight detours by drivers and passengers differently, to accommodate flexibility and mobility accordingly.
\section{Algorithmic Details}

This section covers the algorithm to find all reasonable fits to an offer.
We even solve the more general problem of computing all detours.

For a dataset of $k$ offers $o_i=(s_i,t_i), i=1..k$, and a single request $g=(s',t')$, we want to compute $\mu(s',t')$, $\mu(s_i,s')$ and $\mu(t',t_i)$.
Doing this with $2k+1$ shortest paths queries is one way to do it, but not the fastest.
It is better to adapt an algorithm for distance table computations \cite{ksssw-cmmsp-07}.
This algorithm computes a distance table between a set $S$ of source nodes and a set $T$ of target nodes by performing $\abs{S}$ forward searches and $\abs{T}$ backward searches of an hierarchical routing algorithm, \cite{gssd-chfsh-08} is currently the fastest.
The algorithm works in two phases.
The first phase completes backward searches from each node $t \in T$ and stores $t$ and the distance to $t$ in a \emph{backward bucket} at each reached node.
Because of the hierarchical routing algorithm, the number of reached nodes is very small so only few buckets entries are created.
The second phase executes forward searches from each node $s\in S$, scans the backward buckets at each reached node and updates the distance table with the shortest paths distances found by combining the paths from forward and backward search.
Now assume that $T = \set{t_i \mid i=1..k}$, and we already performed the first phase.
Note that the first phase is independent from $S$.
We can then use a single forward search from $t'$ that scans the backward buckets to compute all $\mu(t',t_i)$.
The problem to compute all the $\mu(s_i,s')$ is solved by switching backward and forward search.
We perform forward searches from all $s_i$ and store $s_i$ and the distance from $s_i$ in a \emph{forward bucket} at each reached node.
A single backward search from $s'$, that scans the forward buckets, can then compute all $\mu(s_i,s')$.

\subsection{Adding and Removing Offers}
To add or remove an offer $o = (s,t)$, we just need to update the forward and backward buckets.
To add the offer, we just perform a forward/backward search from $s/t$ and add the necessary entries to the forward/backward buckets.
To remove the offer, we just perform a forward/backward search from $s/t$ and remove the relevant entries from the forward/backward buckets.

\subsection{Constraints}
In reality, offers and requests have constrains.
They e.g. specify a departure time or time window, have restrictions on smoking or gender, etc.
In this case, we need to extend the definition of an reasonable fit to meet these constraints.
Our algorithm can take advantage of these constraints by filtering offers that clearly violate the constraints.
A prefiltering can be done even during the shortest paths searches from $s'$ and $t'$, e.g. we can safely ignore all offers that depart earlier than the departure time window of the request.
A postfiltering at the end removes the remaining offers with constraint violations.

\section{Experimental Results}
We implemented our algorithm in C++ and tested it against a dataset of real-world ride sharing offers from Germany available on the web. We matched the data against a list of cities, islands, airports and the like, and ended up with about 450 unique places. We tested the data and checked that the length of the journeys are exponentially distributed. This validates assumptions from the field of transportation science. We assumed that requests would follow the same distribution and chose our offers from that data set as well. 

To extend the data set to our approach of arbitrary origin and destination locations, we applied perturbation to the node locations of the data set. For each source node we unpacked the nodes forward search space in the contraction hierarchy up to a distance of 3\,000 seconds of travel time. From that unpacked search space we randomly selected a new starting point. Likewise we unpacked the backward search space of each destination node up to the distance and picked a new destination node. We observed that perturbation preserved the distribution of the original data set.

\begin{figure}[htp]  
	\centering   	
   	\includegraphics[scale=0.48]{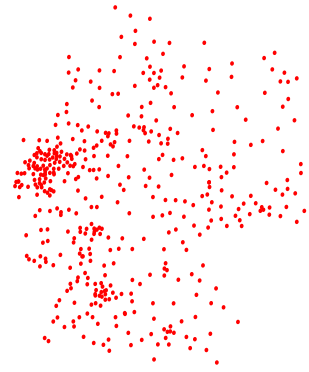}\includegraphics[scale=0.4]{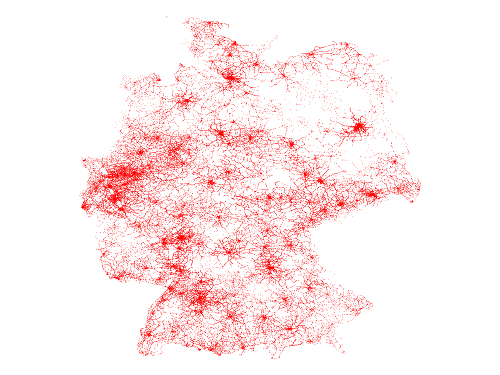}\includegraphics[scale=0.2]{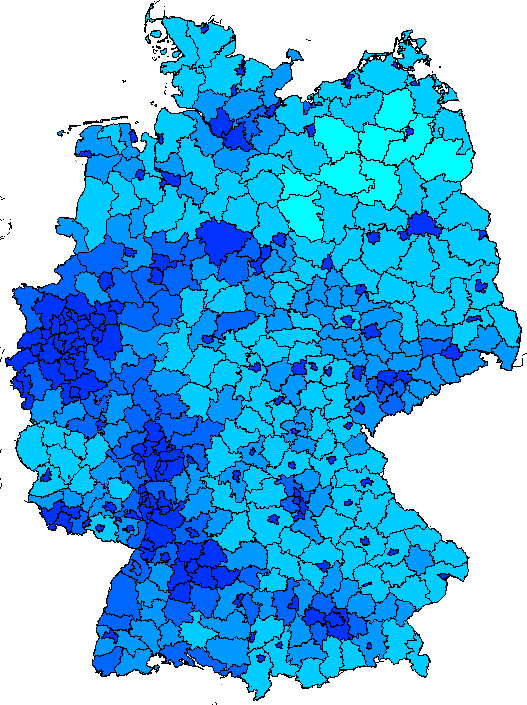}
	\caption{original node locations (left), perturbed node locations (middle), population density (right)}
	\label{fig:nodeplots}
\end{figure}

Figure \ref{fig:nodeplots} compares the original node locations on the left to the result of the node perturbation in the middle. The right side shows a population density plot of Germany\footnote{Picture is an extract of an image available at episcangis.hygiene.uni-wuerzburg.de} to support the validity of the perturbation.

We evaluated our algorithm by measuring the times necessary to match any single request against databases of offers of various sizes.
The numbers in Table~\ref{table:match_times} show that our algorithm can even handle large datasets with 100\,000 real offers within 70\,ms.
In comparison, the fastest speedup technique today, Transit Node Routing \cite{bdsssw-chgds-08,bfss-frrnt-07} requires 1,7\,$\mu$s per query and would take 340\,ms for the largest dataset whereas our algorithm is more than four times faster.
Our algorithm however does not perfectly scale with the number of offers.
The main problem is the bucket size; when it gets too large, our algorithm suffers disproportionally high.
The good news is that when we have a bigger graph then we can handle more offers, since the offers are distributed among more buckets.
Also when the offers are distributed over several days or even weeks, we can split the buckets to ease the problem.

\begin{table}[h]
\centering
\begin{tabular}{l|rrr}
       & 1\,000 & 10\,000 & 100\,000 \\
\hline
random & 1.0 & 5.1 & 63 \\
real   & 0.8 & 4.9 & 70 \\
\end{tabular}
\caption{Request matching time [ms] for different numbers of offers}
\label{table:match_times}
\end{table}

\begin{figure}[htp]  
	\centering   	
   	\includegraphics[scale=0.5]{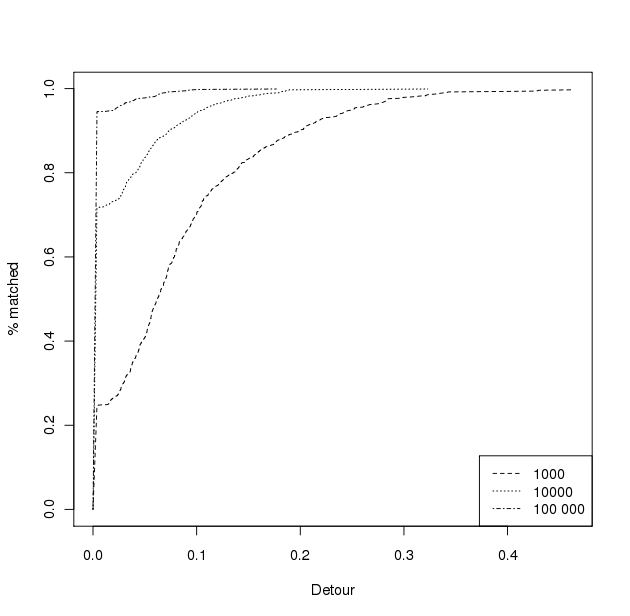}
	\caption{Percentage of rides matched for a given detour}
	\label{fig:matchplot}
\end{figure}

We experimented on the rate with which offers and requests are actually made, but at this point of our research we report on unperturbed data only. Although we had even more encouraging results for perturbed data, we restrict ourselves here for two reasons. First, the modelling of the radial search mechanism of today's ride sharing algorithms is unclear and second, the probability to find a perfect match is close to zero for the perturbed data sets. So, for database sizes of 1\,000, 10\,000 and 100\,000 unperturbed elements we evaluated the fraction of satisfiable requests. Figure \ref{fig:matchplot} and Table \ref{table:match_rate} report on these experiments. We measured the number of requests that could be matched by the maximum allowed detour. For a database size of 10\,000 entries and an maximum allowed detour of 10\% we observed a matching rate of a little more than $94\%$. This is a lot more than the $70\%$ matching rate that we observed without our algorithm.

\begin{table}[h]
\centering
\begin{tabular}{l|rrr}
       & $0.05$ & $0.1$ & $0.2$ \\
\hline
1\,000 & 0.409 & 0.701 & 0.902 \\
10\,000 & 0.836 & 0.941 & 0.997 \\
100\,000 & 0.977 & 0.997 & 1.0
\end{tabular}
\caption{Request matching rate for different values of maximum allowed detour}
\label{table:match_rate}
\end{table}

Note, that the database size of 10\,000 entries is a realistic case and closely resembles the current daily amount of matches made by a known German ride sharing service provider.\footnote{see: http://www.ea-media.net/geschaftsfelder/europealive/geschaftsfelder.html}
\section{Conclusion}
We developed the algorithmic solution to matching ride sharing offers and request for arbitrary starting and destination points. First experiments have shown the validity of the approach. Note, that our algorithm is faster than $2k+1$ distance calculations even with Transit Node Routing. From the observed running times and our experience in static routing we are confident that the algorithm is more than suitable for a large scale web service with potentially hundreds of thousands of users each day.
\bibliographystyle{plain}
\bibliography{references}
\end{document}